\documentstyle[12pt,aasms4]{article}

\received{}
\accepted{ }

\begin{document}
\title {Constraints on the Acceleration of Ultra-High-Energy Cosmic Rays 
in Accretion-Induced Collapse Pulsars} 

\author{Elisabete M. de Gouveia Dal Pino\altaffilmark{1}
\& A. Lazarian\altaffilmark{2} }
\altaffiltext{1}{Instituto Astron\^omico e Geof\'{\i}sico, University
of S\~ao Paulo, Av. Miguel St\'efano, 4200, S\~ao Paulo
04301-904, SP, Brasil;
E-mail: dalpino@iagusp.usp.br,  }
\altaffiltext{2}{Department of  Astronomy,
University of Wisconsin,
Madison, USA; E-mail: lazarian@astro.wisc.edu }

\begin{abstract}
We have recently proposed that  the
ultra-high energy  cosmic rays (UHECRs) observed above the GZK
limit could be mostly protons accelerated in reconnection sites just above the
magnetosphere of newborn millisecond pulsars originated by accretion
induced collapse (AIC-pulsars).
Although the expected rate of AIC sources in our own Galaxy
is very small ($\sim \,  10^{-5}$ yr$^{-1}$),
our estimates have shown  that the observed total
flux of UHECRs could be
obtained  from the integrated contribution from  AIC-pulsars
of the whole distribution of galaxies located within a distance which
is unaffected by the GZK cutoff ($\sim \, 50 $ Mpc).
We presently examine
the potential acceleration mechanisms  in the magnetic reconnection site and
find that first-order Fermi acceleration  cannot provide 
sufficient efficiency. 
To prevent synchrotron
losses, only very small
deflection angles of the UHECRs would be allowed in the strong
magnetic fields of the pulsar,
which is contrary to the requirements for efficient Fermi acceleration.
This leaves the one-shot acceleration via an induced electric
field within the reconnection
region as the only viable process for UHECR acceleration.
We formulate the constraints on both the magnetic field topology
and strength in order to accelerate the particles and allow them
to freely escape from the system. Under fast reconnection
conditions, we find that AIC-pulsars with surface magnetic fields
 $10^{12} $ G  $<  \, B_{\star} \lesssim  $ $10^{15}$ G
and spin periods
1 ms $\lesssim \, P_{\star} \, < \, $ 60  ms,
are able to accelerate particles to energies
$\geq \, 10^{20} $ eV, but the magnetic field just
above the Alfv\'en surface must be predominantly toroidal for the
particles to be allowed to escape from the acceleration zone
without being deflected.
Synchrotron losses
bring potentially important constraints
on the magnetic field geometry of $any$ UHECR accelerators
involving  compact
sources with strong magnetic fields.
\end{abstract}


\section{Introduction}
The origin and nature of the observed cosmic ray events with energies beyond
10$^{20}$eV
remains a mystery. Up to now, more than 55 with $E >
4 \times 10^{19}$ eV  have
been detected (Takeda et al. 1999), and  the similarities of
their air showers  with those of cosmic rays at lower
energies suggest that they could be mostly protons
(Protheroe
1999), although the possibility that they are heavy nuclei, $\gamma -
$rays,
neutrinos, or exotic particles cannot be totally
ruled out at the present. If these ultra-high energy cosmic rays (UHECRs) are
mostly protons, then they should be
affected
by
the expected Greisen-Zatsepin-Kuzmin (GZK) energy cutoff ($\sim 5\times
10^{19}$ eV for protons), due photo-pion production by interactions
with
the
cosmic microwave background radiation, unless they are originated at
distances closer than about 50 Mpc  (e.g., Protheroe \& Johnson 1995,
Medina Tanco, de Gouveia Dal Pino \& Horvath 1997). On the other hand,
if
the they are protons from nearby sources
(located within $\sim $
 50 Mpc), then they should be little deflected by the
intergalactic and Galactic magnetic fields and point toward
their sources (e.g., Stanev 1997, Medina
Tanco,
de Gouveia Dal Pino \& Horvath 1998).

Magnetic confinement of UHECR by the Galaxy is very
difficult since the Larmour radius for a proton with energy $E \simeq 10^{21}$
eV is of the order of few $100$ kpc, and
the present data although statistically modest, also seem to indicate an
extragalactic origin for the UHECR events. There is no significant
large-scale anisotropy related to the Galactic disk, halo,
or
the local distribution of galaxies, although some clusters of events
seem
to
point to the supergalactic plane (Takeda et al. 1999).


In an attempt to disentangle the puzzle presented by the detection of these
particles with extremely high energies, several  source candidates and
acceleration mechanisms have been
invoked, but all of them have their limitations
(see, e.g., Protheroe 1999, Blandford 2000, and Olinto 2000, 2001, for reviews).
The proposed models can be separated in two classes: (i) the so called
astrophysical Zevatrons, or $buttom-up$ models, which involve searching for
acceleration sites in known astrophysical objects that can reach the required
ZeV energies, such as compact sources with high magnetic fields and rotation
rates, or powerful shocks; and (ii) the $top-down$ models that involve the decay
of high mass relics from the early Universe.
Among the Zevatrons, shock acceleration
in the radio lobes and jets of powerful radio sources and AGNs may
appear
as an
attractive possibility (e.g., Rachen \& Biermann 1993), but the lack of
direct
correlation of the arrival directions of most of the observed UHECRs
events
with nearby radio galaxies or AGNs poses some difficulties to these
candidates. Among the top-down models, the production of UHECRs by hadronization
of quarks and
glouns
generated during the evaporation of primordial black holes located
mainly
at
an extended Galactic halo has been recently examined as an
alternative
possibility (Barrau 1999).

Other potential Zevatrons are unipolar inductors, like millisecond pulsars
with very strong magnetic fields (B $>10^{12} $ G), or magnetars.
Particles can,
in principle, extract the required energies from an induced e.m.f. across the
few open field lines of a
rapidly
rotating magnetar, although a large electric field parallel to the magnetic
field can be easily shorted by electron-positron pairs. Alternatively,
the particles can be accelerated in reconnection
sites of magnetic loops produced, e.g.,
by Parker instability,
on the surface of a magnetar (Medina Tanco, de Gouveia Dal Pino \&
Horvath
1997), but the accelerated particles will probably lose
most of their energy gain by curvature radiation while dragged along by
the magnetic dipole  field (Sorrell 1987).

To overcome these difficulties imposed by acceleration in regions located close
to the surface of a pulsar, in a recent work (de Gouveia Dal Pino \& Lazarian
2000, hereafter Paper I), we
have speculated that UHECRs could be mostly protons
accelerated in
magnetic reconnection sites $outside$ the magnetosphere of  newborn
millisecond pulsars produced by accretion induced collapse (AIC) of a
white dwarf (e.g.,
 Woosley \& Baron 1992 and references therein).
As stressed in Paper I, the
accretion flow spins up the star and confines the magnetosphere
to a radius $R_X$ where  both plasma stress in the accretion disk,  and
magnetic
stress
balance
(Arons 1993). At this radius, which also defines the inner radius of
the
accretion disk,   the equatorial flow diverts into a funnel inflow
along the closed
field-lines toward the star, and
a centrifugally
driven wind outflow (Gosh \& Lamb 1978, Arons 1986,
Shu et al. 1999).
To mediate the geometry of dipole-like field lines of the star with
those
opened by the wind and those trapped by the funnel inflow emanating
from
the $R_X$ region,
a surface of null poloidal field lines is required.
It is labeled as "$helmet$ $streamer$"
in
Figure 1a.
(see also Fig. 1 of Paper I).

Across the  null surface, the poloidal field suffers a sharp reversal
of direction. According to the Amp\`ere's law, large electric currents
must flow out of the plane shown in Figure 1a, along the null
surfaces, and
in the presence of finite electric resistivity, dissipation of these
currents will lead to reconnection of the oppositely  directed
field lines (e.g., Biskamp 1997, Vishniac \& Lazarian 1998).
Helmet streamers (or flare loops) are also present in the magnetic
field configuration
 of the
solar corona.
The magnetic
energy released by reconnection in the helmet streamer
 drives violent
outward motions in the surrounding plasma
that accelerate copious
amounts
of solar cosmic rays without producing many photons (Reames 1995).
A similar process may take place in the helmet streamer of
young born AIC-pulsars.

The particular mechanism of particle acceleration
during reconnection events is still unclear in spite of
numerous attempts
to solve the problem
(see
LaRosa et al 1996,  Litvinenko 1996).
Cosmic rays from the Sun confirm that the process is sufficiently
efficient in spite of the apparent theoretical difficulties for its
explanation. In this situation we attempt to place constraints
on the geometry of the magnetic field at the reconnection sites and
identify the most promissing processes that can provide UHECR
acceleration at AIC-pulsars.

We have argued in Paper I   that
 protons could be accelerated to the ultra high energies in AIC-pulsars by the
large induced electric field
within the reconnection region
(e.g.,
Haswell, Tajima, \& Sakai 1992, Litvinenko 1996).
This allowed us to obtain a flux of
UHECRs which was consistent with the observations, but
the physics of the acceleration process itself was not discussed there.
In this paper, we investigate  the potential acceleration mechanisms in the
reconnection site. We first examine the conditions at which the rate of magnetic
reconnection itself can be maximized in the presence of anomalous resistivity
(\P 2). Then, we show that
first-order Fermi acceleration in the reconnection site is possible, but
it cannot produce UHECRs because of the synchrotron
losses that the particles experience in the strong magnetic fields of the
pulsars (\P 3). We finally formulate the necessary constraints  on the
magnetic
field geometry in order to enable one-shot acceleration by the induced electric
field
in the reconnection region, re-evaluate the expected UHECR spectrum and flux
(\P 4), and conclude with a brief discussion of the implications of our results
(\P 5).

\section{Anomalous Resisistivity  in the Reconnection Region}
The commonly referred problem in classical reconnection schemes is that they use
to provide too
slow reconnection rates. For instance, Sweet-Parker reconnection implies a
reconnection rate that is smaller than the Alfv\'en speed by
a factor $ R_{eM}^{-1/2}$, where $R_{eM}$ is the magnetic Reynolds number.
This factor can be as small as
$10^{-10}$ in the ISM, and makes the reconnection
rate extremely slow. Very slow reconnection rates would make the losses too
large and the acceleration inefficient.
However, there is ample observational evidence
that reconnection in Astrophysics takes place at much larger rates which are
instead comparable with
the Alfv\'en speed (e.g., Dere 1996). In a
 recently suggested
 model of turbulent reconnection, Lazarian \& Vishniac (1999, hereafter LV99;
2000) have  appealed
to an inevitable wandering of magnetic field lines as the ultimate
cause of fast reconnection\footnote{Recently Kim and Diamond (2001)
noticed that in the
problem of reduced dimensionality (less than 3D),
the reconnection is slow in spite of
the stochastic character of the magnetic field lines. This is consistent with
the claim in Lazarian \& Vishniac (1999), where the three-dimensionality
of the field wandering was required.}.
The scheme explains naturally flaring and
other Astrphysical phenomena, but numerical testing is still
required. A competing model based on plasma properties (Bisckamp, Schwarz
\& Drake, 1997)
to stabilize the Petcheck reconnection layer has also been recently
put forward.
Further research should determine the domains of each particular
scheme's applicability.

Presently, let us focus our attention on regions of strong
magnetic fields, like
those around pulsars. In this case, fast reconnection
(with $v_{rec} \sim v_A$) is ensured by
anomalous resistivity (see Parker 1979).
Following LV99, we can estimate the width of the current sheet
for which the resisitivity should be anomalous:
\begin{equation}
\delta=\frac{c\Delta B}{4\pi n Z e u}
\end{equation}
where $\Delta B \simeq B$ denotes the change of the magnetic field across
the reconnection region, $n$ the particle number density,
and $u$ the thermal velocity of the ions of charge $Ze$.
At the radius $R_X$, where the magnetic field corrotates with the pulsar (see
Fig. 1a), the particle density has approximately the Goldreich-Julian  value
(e.g., Goldreich \& Julian 1967)
$n \simeq B(R_X) \Omega_{\star}/4 \pi Z e c$,
which implies
\begin{equation}
\delta \simeq 10^9 {\rm cm} \, \Omega_{2.5 k}^{-1} u_{9}^{-1}
\end{equation}
where $u_{9}$ is the thermal velocity in units of $10^{9}$ cm s$^{-1}$, and
$\Omega_{2.5k} = \Omega_{\star}/ 2.5 \times 10^3$ s$^{-1}$, with
$\Omega_{\star}$ being the angular speed at the stellar surface. Although
this estimate is somewhat crude, the value of
$\delta \simeq 10^9$ cm indicates that the conditions in the pulsar are more
than appropriate to produce fast reconnection
through anomalous resistivity over the entire region since
$\delta > > R_X$ (see below).

Therefore, whatever processes are invoked to accelerate the particles,
it is realistic to assume that the reconnection velocity
that we deal with is an appreciable fraction of the local Alfv\'en
velocity.
Besides, as in the conditions we deal with this speed
approaches $c$, the expected acceleration rate can be very large.
Any limitations on the efficiency at which acceleration may occur will then come
solely from the way by which particles will manage  to
escape from the strong magnetic fields around the pulsar.

\section{Synchrotron Losses} 

A schematic representation of a reconnection region is shown in Figure ~1b. The
upper and
lower parts of the magnetic flux move towards each other with a velocity
$v_{rec}$. As a result, charged particles in the upper 
part of the reconnection
zone "see" the lower part of the magnetic flux to approach them with
a velocity $2v_{rec}$, and an  acceleration process analogous
to the first-order Fermi acceleration of cosmic rays in magnetized shocks
(e.g., Longair 1992) may take place.
In fact, it is straightforward to show that, during  a round trip
between the upper and lower magnetic
fluxes, a particle will suffer  an average increment of energy
\begin{equation}
\langle\frac{\Delta E}{E}\rangle
\simeq \frac{8}{3}\frac{v_{rec}}{c},
\end{equation}
\noindent
so that after a number of crossings the particles could in principle reach 
the required ultra-high energies.
However, in the very strong magnetic fields present in the young pulsar,
the particle energy  losses due to synchrotron radiation can be rather substantial.
 Let us
evaluate them.

A well known formula for the power radiated by a
 moving charged particle is (e.g., Jackson 1981)
\begin{equation}
P_r= \frac{2}{3}\frac{e^2}{m^2 c^3}\left[\left(\frac{d{\bf p}}{d\tau}\right)^2
-\frac{1}{c}\left(\frac{dE}{d\tau}\right)^2\right]
\label{power}
\end{equation}
where $d\tau=dt/\gamma$ is the proper time element. A usual assumption
in the derivation of the synchrotron losses formulae is that
$c \, d{\bf p}/d\tau\gg dE/d\tau$, which means that the particle momentum,
${\bf p}$, changes rapidly in direction without relevant changes in E during a
particle revolution around the magnetic field line. In this case, after simple
algebra one can easily get
\begin{equation}
\frac{\delta E}{E}\approx 3.7 \times 10^{12} \, \delta\varphi B_{13} E_{20}^2
\sin\theta
\label{dev}
\end{equation}
where 
 $\delta E$ is the amount of energy lost by the particle of energy
$E_{20}=E/10^{20}eV$,  when deflected by an angle $\delta \varphi$ in a magnetic
field
$B_{13}=B/10^{13}G$,
and $\theta$ is the pitch angle between the magnetic field
and the particle velocity.

It is clear from Eq.(4), however, that for the regime that we are
interested,
i.e., in the case of UHECR acceleration, the energy losses per revolution period may
be very
substantial. In this case, the classical synchrotron loss formulae should
be modified. Rewriting Eq.(4) in the form
\begin{equation}
\frac{P_r}{2e^2/3 m^2c^3}=\left(\frac{d{\bf p}}{d\tau}\right)^2-\frac{1}{c^2}
\left(\frac{dE}{d\tau}\right)^2
\label{inter}
\end{equation}
and taking into account that $P_r=dE/dt=dE/d\tau$ is a Lorentz
invariant we get a quadratic equation for $P_r$:
\begin{equation}
P_r/D=(\gamma \omega_B E/c)^2 - 1/c^2 P_r^2,
\end{equation}
where $D=2e^2/3 m^2c^3$, and $\omega_B=eB\sin\theta/mc\gamma$ is the Larmor
frequency of the particle.
The solution of this equation is
\begin{equation}
P_r=\frac{c^2 [-1+(1+\Delta)^{1/2}]}{2D}
\end{equation}
where $\Delta=(4/c^2)(D(\gamma \omega_B E/c)^2$. In the case of
$\Delta\gg 1$, it reduces  to
\begin{equation}
P_r \, \simeq \, \gamma \omega_B E
\end{equation}
which means that for $\Delta\gg 1$, the energy loss
per deflection $\delta \varphi$ will be
$\delta E \simeq P_r \delta \varphi/ \omega_B$, or
\begin{equation}
\frac{\delta E}{E}\approx 1.04\times 10^{11} \delta \varphi E_{20} \sin \theta,
\label{loss2}
\end{equation}
which is independent of the magnetic field intensity. Comparing Eqs. (10) and
 (5), we see that the corrected synchrotron losses for UHE particles
decreases by only an order of magnitude relative to the original evaluation.
Therefore, in the UHE particle regime, the synchrotron losses in strong magnetic
fields are so large, even for very small deflection angles, that they exclude
any possibility for particle acceleration via reflecting particles back and forth
within the reconnection zone, as required in first (or second) order Fermi
acceleration processes.
Fermi acceleration would be possible for UHECRs only if the
magnetic fields in the system were relatively weak. As indicated by Eq. (5), for
an $E_{20}$ particle, $B$ must be $\lesssim 1$ G in order to cause no relevant
energy losses for finite particle deflection angles. Since in the AIC-pulsars,
very strong magnetic fields are present and actually provide the energy
reservoir for particle acceleration,
we conclude that Fermi processes are not suitable in this case.  We show
below instead, that
direct acceleration by the induced electric field  in the reconnection zone is
the only viable mechanism.

\section{Model for One-Shot UHECR Acceleration in the Reconnection Region}

In Paper I, we have argued that the protons could be accelerated by a large
induced electric field within the reconnection region in the helmet streamer
(Haswell et al. 1992, Litvinenko 1996), but now we find that an additional
constraint should be  applied to the magnetic field  geometry, namely, that the
accelerated
UHECRs should escape without experiencing any significant deflection, otherwise
their gained energy would be lost by synchrotron effects in the strong poloidal
fields.

In the idealized two-dimensional field topology suggested in Paper I (see Fig.
1a), the reconnection site was a simple straight sheet along which the particles
were accelerated by an electric field coming out of the plane shown in Fig. 1a,
so that the particles could escape freely from the system on the normal
direction to the plane of Fig. 1a. However, in a more realistic
three-dimensional geometry, the reconnection region of Fig. 1a actually
describes a
cone around the rotation axis of the star, so that the accelerated particles
escaping perpendicularly out of the plane of Fig. 1a will be eventually strongly
deflected by poloidal field lines in their way out of the system, thus loosing
most of their energy (Eqs. 5 and 10).

On the other hand, it is well known from
magnetized-winds theory (e.g., Spruit 1996) that beyond the Alfv\'en surface
(that is the surface at which the wind flow emerging from the disk/star system
reaches the Alfv\'en velocity, $v_A$), the inertia of the gas causes it to lag
behind the rotation of the field line, so that the poloidal lines ($\vec B_p$)
get wound up thus developing a toroidal component ($\vec B_{\phi}$) (see a
schematic representation of the field line winding process in Figure 2a). In the
extreme relativistic limit, the inertial forces in the flow are so high near the
Alfv\'en surface that they bend the poloidal lines into nearly horizontal shape,
and the field becomes preferentially spiral, i.e. $B_{\phi}\gg B_{p}$ (e.g.,
Camezind 1987). In this case, as the accelerating electric field, which is
perpendicular to the annihilating
magnetic field lines, is mostly poloidal ($\vec \epsilon_p$) (see Figure 2b), the
particles will be accelerated (and allowed to escape) along the poloidal
direction without being deflected by the field lines, therefore, without
suffering substantial synchrotron losses.

Following Paper I, we adopt Shu et al. (1999)
field geometry
of  magnetized stars accreting matter from a disk (Fig. 1a), but take into
account the formation of a toroidal magnetic field above the Alfv\'en
surface (Fig. 2a).

For a Keplerian disk, the inner disk edge $R_X$ rotates
at
an angular speed
($G M_{\star}/R_X^3)^{1/2}$, and  equilibrium between gravity
and centrifugal
force at $R_X$ will lead to co-rotation of the star
with the inner disk edge, i.e.,
$R_X \, = \, (G M_{\star}/\Omega _{\star}^2)^{1/3}$,
which for typical millisecond pulsars with
rotation periods
$P_{\star} = 2\pi/\Omega_{\star} \simeq 1.5 - 10$ ms, mass
$M_{\star} \, \simeq  \, $1 M$_{\odot}$, and  radius
$R_{\star} = 10^6$ cm, gives
$R_X \simeq (2   \, - 7) \times 10^{6}$ cm $R_{6}$,
where $R_{6} = R_{\star}/ 10^6$ cm.
The magnetic field intensity in the $R_X$ region is given by (Arons 93, Paper
I)
\begin{equation}
B_X \, \simeq \, B_{dipole}(R_X) \left( {\frac{R_X}
{ \Delta R_X} }\right)^{1/2}\,
\label{BX}
\end{equation}
\noindent
where,
$\Delta R_X$ is the width of the reconnection zone,
$B_{dipole}(R_X) = B_{\star} \, (R_{\star}/R_X)^{3}$ is the
magnetic field that would be
present in the absence of the shielding disk, and
$B_{\star}$ is the magnetic field at the surface
of the star.

In a  reconnection event, an electric
field arises inductively along the poloidal direction
because of plasma flow across the $\vec B$ lines.
The condition that particles of charge $Ze$  can be
accelerated to energies $E$ by an electric voltage drop, $V$, is given by
$E=Ze V = ZeB_{\phi}\xi L_X$, where
$\xi =v_{rec}/v_A \simeq v_{rec}/c$
is the reconnection efficiency factor,
and $L_x$ is the length of the reconnection region (see Fig 2b).
For fast
reconnection, $\Delta R_X/R_X \lesssim 1$
\footnote{This condition on $\Delta R_X$, which is naturally satisfied if the
reconnection is fast (or the resistivity is anomalous; see \P 2),  further
ensures that the thickness of the
neutral reconnection zone is large enough to allow the particles to move
freely in the poloidal direction without being deflected by the field lines
($B_\phi$) while escaping from the system.
},  and $L_X/\Delta R_X\approx v_A/
\xi v_A \approx \xi^{-1}$ (e.g., LV99). This results   $E \approx Ze \Delta R_X B_{\phi}$,
or
\begin{equation} \frac{E}{R_X B_{\phi} Ze}\approx \frac{\Delta R_X}{R_X}
\lesssim 1 \label{5}
\end{equation}
Assuming that $B_{\phi}$ is of the order of $B_X$ we can get a limit
on $B_{\phi}$ using
Eq. (12),
and substituting
$R_X=(GM)^{1/3}/\Omega^{2/3}$ for the corrotation radius, namely
\begin{equation}
B_{\phi} \, \gtrsim \,\frac{E}{R_X Ze} \approx \frac{E\Omega^{2/3}}{(GM)^{1/3}
Ze}.
\end{equation}
Using
Eq. (11), it is easy to obtain
\begin{equation}
B_{13}  \gtrsim  0.8\times E_{20} Z^{-1} \Omega^{-4/3}_{2.5k}
\label{Bstar}
\end{equation}
\noindent
where we have assumed
$M_{\star} \, =  \, 1  M_{\odot}$,
$R_{\star} = R_6$,
$E_{20} \, = \, E/10^{20} $ eV,
$\Omega_{2.5k} = \Omega_{\star}/ 2.5 \times 10^3$ s$^{-1}$,
and $B_{13} = B_{\star}/10^{13}$ G. We note that a similar condition was derived
in Paper I, but  from a
different requirement, i.e.,  that $\Delta R_X$ should be larger than
about twice the particles Larmour  radius.

Eq. (14) (see also Fig. 2 of Paper I) indicates that
stellar magnetic fields
 $10^{12} $ G $ <  B_{\star} \lesssim  $ $10^{15}$ G
and angular speeds
$4 \times 10^{3}$ s$^{-1}$ $\gtrsim \Omega_{\star} \, > \,  10^{2} $
s$^{-1}$,
which correspond to spin periods
1 ms $\lesssim \, P_{\star} \, < \, $ 60  ms, are able to accelerate
particles to energies $E_{20} \,  \gtrsim $ 1. Slower pulsars with angular
frequencies $\Omega_{\star} \lesssim 10^{2} $ s$^{-1}$
require too large surface magnetic fields ($B_{\star} > \, 10^{15}$
G)
to efficiently accelerate the
particles.
The values above are perfectly compatible with the parameters
of young pulsars and
Eq. (14) is thus a good representation of the typical conditions
required
for particle acceleration in reconnection zones of AIC-pulsars through
one-shot process.

A newborn millisecond pulsar spins down due to
magnetic
dipole radiation in a time scale given by
$\tau_{\star} = \Omega_{\star}/\dot \Omega_{\star} \simeq
\left({\frac{ I c^3 }{B_{\star}^2 R_{\star}^6 \Omega_{\star}^2 }
}\right)$,
which for a moment of inertia
$I = 10^{45}$ g cm$^2$
gives
$\tau_{\star} \simeq 4.3 \times 10^7 $ s $  B_{13}^{-2} \,
\Omega_{2.5k}^{-2}$.
We have shown in Paper I that the
condition that the magnetosphere and
the disk stresses are in equilibrium
at
the inner disk edge results a disk mass accretion rate
\begin{equation}
\dot M_D \, \simeq \,  3 \times 10^{-8} M_{\odot} {\rm s}^{-1} \, {\alpha_2}^{-
2} \, B_{13}^2 \, \Omega_{2.5k}^{7/3}
\end{equation}
\noindent
where
$1 \gtrsim \alpha_2 > 0.5$
 measures the amount of
magnetic dipole flux that has been pushed by the disk accretion flow to
the inner edge of the disk
(Gosh \& Lamb 1978,
 Shu et al. 1994).
$\dot M_D $ is obviously
 much larger than the
Eddington accretion rate,
$\dot M_{Edd} \simeq 7.0 \times 10^{-17} M_{\odot} $ s$^{-1}
(M_{\star}/ M_{\odot})$, but this supercritical accretion will
last for a time  $\tau_{D}$, which is only
a small fraction ($f_D$) of  $\tau_{\star}$ (Paper I).
Considering that advection dominated inflow-outflow solutions involving
supercritical accretion onto neutron stars predict a
total mass
deposition on  the star
$M \sim $  few $0.01 M_{\odot}$ (e.g., Brown et al. 1999),
we have found that
$\tau_D \simeq M/\dot M_D \simeq 1.3 \times 10^6 $ s, or
 $f_D =\tau_D /\tau_{\star} \simeq 0.03$
(Paper I),
which implies that
the most violent reconnection events can live
at least for several days.
As the acceleration of UHECRs
in the reconnection zone
 will occur during 
the supercritical accretion event,
the spectrum evolution of
the
accelerated UHECRs will be determined by
$\tau_D = f_D \tau_{\star}$
(see below).

The rate of magnetic energy that can be extracted from the
reconnection region is
$\dot W_B \simeq (B_X^2/ 8\pi) \, \xi v_A \, (4\pi R_X \,L_X
)$,
where $v_A \sim \, c$.
Substituting the previous relations into this equation, one finds
\begin{equation}
\dot W_B \, \lesssim \,  2.6 \times 10^{46}   {\rm erg  s}^{-1} \, \xi
\, B_{13}^2 \, \Omega_{2.5k}^{8/3}.
\end{equation}
\noindent
According to our model assumptions, in a  reconnection event an electric
field arises inductively
and
once a particle decouples from the injected fluid, it
will be ballistically accelerated by this field.
 The UHECR flux emerging from the reconnection site can then be
estimated as
\begin{equation}
\dot N \, \simeq \, {\frac {\dot W_B } { E}}  \, <\, 1.6 \times 10^{38}
\, {\rm s}^{-1} \, B_{13}^2 \, \Omega_{2.5k}^{8/3} \,E_{20}^{-1}
\end{equation}
\noindent
for particles with energy $E \gtrsim  10^{20} $ eV, and
 the particle spectrum $N(E)$ is obtained from
$\dot N \, = \, N(E) \, {\frac {dE } { dt}}  \,
\simeq \, N(E) \,  {\frac {dE } { d \Omega_{\star} } }  \, \dot
\Omega_{\star}/f_D$,
or
\begin{equation}
N(E) \, \lesssim
\, 1.6 \times 10^{33} \, {\rm GeV}^{-1} \,   Z^{-1/2} \,
B_{13}^{-
1/2} \, E_{20}^{-3/2}
\end{equation}
\noindent
where Eq. (13)
 has
given
$ d \Omega_{\star}/dE  \simeq 1.2 \times 10^{-5}  {\rm erg^{-1} \rm s^{-1} } \, Z^{-3/4}
\,  E_{20}^{-1/4} \, B_{13}^{-3/4} $
 (with the signal made equal in Eq. 13).
Eq. (18) above is again similar to the one obtained in Paper I, and predicts
that  $N(E) \propto E^{-3/2} = E^{-1.5}$, which is a flat spectrum
in
good
agreement with observations (e.g., Olinto 2000).


As the total number of sources formed via AICs
in our Galaxy is limited by nucleosynthesis constraints to a very small
rate  ${\tau_{AIC} }^{-1} \, \simeq  \, 10^{-5}$ yr$^{-1}$
(Fryer et al. 1999),
we find that the probability of having UHECR events produced solely in the
Galaxy is very small (Paper I). However, we can
 evaluate the integrated
contribution due to  AICs from all the galaxies located within a
volume
which is not affected by the GZK effect, i.e., within a radius
$R_{50} = R_G/ 50$ Mpc.
Assuming  that each galaxy has essentially the
same rate of AICs as our own Galaxy and taking the standard galaxy
distribution
$n_G \simeq \, 0.01 \, e^{\pm 0.4}\, h^3 $ Mpc$^{-3}$ (Peebles 1993)
(with the Hubble parameter defined as $H_o = h$ 100 km s$^{-1} $
Mpc$^{-1}$), the resulting flux at $E_{20} \, \geq $  1 is
$F(E) \, \simeq \, \,  N(E) \, n_G \, {\tau_{AIC}}^{-1} \, R_{G}$
(Paper I),
which gives
\begin{equation}
F(E) \, \lesssim \, 3.1 \times 10^{-29} \, {\rm GeV}^{-1}
{\rm cm}^{-2} {\rm s}^{-1} \,Z^{-1/2} \, B_{13}^{-1/2} \,
E_{20}^{-3/2} \, {\tau_{AIC,5}}^{-1} \, n_{0.01}  \,
R_{50}
\end{equation}
\noindent
where ${\tau_{AIC,5}}^{-1} \, = \, {\tau_{AIC}}^{-1}/ 10^{-5}$ yr$^{-
1}$,
 and
$n_{0.01} = n_G/0.01 $ h$^3$ Mpc$^{-3}$.
Observed data by the AGASA experiment (Takeda et al. 1999) gives a flux
at
$E = $10$^{20}$  eV  of
$F(E) \, \simeq  \,  4 \times \, 10^{-30}$ Gev$^{-1}$ cm$^{-2}$ s$^{-
1}$, so
that the efficiency of converting magnetic energy into UHECR should be
$ F(E)_{obs}/F(E)/ \simeq \xi^{\prime} \, \gtrsim \,
 0.1$
in order to reproduce such a signal.
%

\section{Conclusions and Discussion}

In this work, we have studied the processes that could accelerate protons in the
reconnection sites just above the magnetosphere of very young millisecond
pulsars originated by accretion-induced collapse.
Our calculation of the synchrotron losses
have testified that any tangible deflection of UHECR in a strong
magnetic field results in unrecoverable energy losses. Although
this does not prevent the UHECRs to be deflected by weak magnetic
fields, like, e.g., the Galactic magnetic field, it makes it essential that  the
propagation
of the UHECRs in the strong magnetic fields of the pulsar be straight. This
finding
not only imposes limits on the geometry of the reconnecting magnetic
field flux in our model, but also brings potentially important constraints
on the magnetic field geometry of any accelerating models involving compact
sources with strong magnetic fields, like for instance, the pulsar wind model
proposed by Blasi, Epstein, \& Olinto (2000) to generate UHECRs.


Although the back and forth bouncing of protons within the reconnection
region in a Fermi-like mechanism would entail prohibitive energy losses in the
case of UHECRs,
the process may be still important for Solar physics, where proton
acceleration happens within flares.

On the other hand, our model
allows one-shot acceleration of the particles by the electric
field created in the reconnection region, but the magnetic field just
above the Alfv\'en surface must be predominantly toroidal for the particles to be allowed to escape
freely along the poloidal direction in the acceleration zone without being deflected
by the magnetic field lines.
Under fast reconnection conditions, we find that AIC-pulsars with
surface magnetic fields
$10^{12} $ G  $<  \, B_{\star} \lesssim  $ $10^{15}$ G
and spin periods
1 ms $\lesssim \, P_{\star} \, < \, $ 60  ms,
are able to accelerate
particles to energies $\geq \, 10^{20} $ eV. These limits can be
summarized by the condition
$B_{\star} \,  \gtrsim \, 10^{13}$ G $  (P_{\star}/2.5 $ ms)$^{4/3}$  which is
valid for
fiducial stellar and accretion disk/reconnection parameters. The produced
particle spectrum is very flat, as required by the data,
and the total flux is
given  by the integrated contribution from  AIC-pulsars of the
whole distribution of galaxies within the local universe.
However, the efficiency factor for converting magnetic energy in the
reconnection to acceleration of the UHECRs needs to be $ \xi\prime \, \gtrsim \,
0.1$
in order to reproduce the observed flux.
These results predict, therefore,
an extragalactic origin of the UHECRs, in agreement
with the present observations, but as data collection improves, we
should expect some sign of correlation with the local distribution of
galaxies.

In Paper I, we have formulated some constraints on the size
of the
acceleration
zone due to energy losses by pion and
and $e^{\pm}$
pair production
off interactions with an ($overestimated$) population of photons
from the radiation field produced
by the accretion disk.
The same constraints could, in principle, be applied also in  the present
analysis. Although they do not seem to be important  enough to prevent
UHECR production, we should  note
that, as in Paper I, the lack of  a
theory  that can precisely determine  the amount of
photons effectively reaching the reconnection zone well beyond the disk
has made the estimate above very uncertain. In the future, a self-consistent
calculation of the evolution of the star/disk system during the
supercritical phase, involving advected-dominated
inflow-outflow solutions, is required.

Finally, we should note that although we have essentially discussed the
acceleration of protons in the AICs, Eqs. (14), (18) and (19) indicate that the
proposed mechanism could be, in principle, also applicable to heavier nuclei
(e.g., Fe, for which $Z = $ 26). However, since most of the UHECR
events  from AICs must come from extragalactic sources,  it would be
more difficult to propagate the nuclei heavier than the protons, because of the
additional photonuclear disintegration they suffer (Elbert \& Sommers
1995).

\acknowledgements
We thank the anonymous referee for his(her) comments on this work.
E.M.G.D.P. has been partially supported by grants of the Brazilian
Agencies FAPESP and CNPq. A.L. has been also partially supported by FAPESP
(grants 2000/12136-4, and 1997/13084-3).
A.L. thanks the Astronomy Department of IAG-USP for their hospitality.

\newpage
{\bf Figure Caption}

\noindent
Figure 1. a) Schematic drawing of the magnetic field geometry and the gas
accretion flow in the inner disk edge at $R_X$. UHECRs are accelerated in
the magnetic reconnection site at the helmet streamer (see text). The
figure also indicates that coronal winds from the star and the disk
help the magnetocentrifugally driven wind at $R_X$ to open the field
lines
around the helmet streamer (extracted from Paper I); b) Schematic representation
of a reconnection region.

\noindent
Figure 2. a) Development of a toroidal field, $B_{\phi}$, from the winding up of
a poloidal line, $(B_{p})$, above the Alfv\'en surface. With each rotation of the
line, a loop of field is added to the flow at the Alfv\'en surface; b) Schematic
representation of the reconnection region just above the Alfv\'en surface.


\begin{thebibliography}{}

\bibitem[] { }
Aly, J. 1980, A\&A, 86, 192

\bibitem[] { }
Arons, J.  1993, \apj, 408, 160

\bibitem[] { }
Arons, J. 1986, in Plasma Penetration into Magnetospheres, eds. N. Kylafis, J.
Papamastorakis, and J. Ventura (Iraklion: Crete Univ. Press), 115


\bibitem[] { }
Barrau, A. 1999, Astroparticle Phys., in press


\bibitem[] { }
Bird, D.J. et al. 1995, \apj, 441, 144

\bibitem[] { }
Biskamp, D., Schwarz, E., \& Drake, J.F. 1997, Phys. Plasmas, 4, 1002

\bibitem[] { }
Biskamp, D. 1997, in Advanced Topics in Astrophysical and Space Plasmas, eds.
E.M. de Gouveia Dal Pino, A. Perat, G.A. Medina Tanco, and A.C.L. Chian (The
Netherlands: Kluwer), 165

\bibitem[] { }
Blandford, R. 2000, in Particle Physics and the Universe. Ed. Bergstrom, 
Carlson and Fransson, Phys. Scripta, T85, 191

\bibitem[] { }
Blasi, P., Epstein, R.I., \& Olinto, A.V. 2000, \apj, 533, L123

\bibitem[] { }
Camezind, M. 1987, A\&A, 184, 341

\bibitem[] { }
Cronin, J.W. 1999, Rev. Mod. Phys., 71, 165


\bibitem[] { }
de Gouveia Dal Pino, E.M. \& Lazarian, A.  2000, \apj, L31 (Paper I)
                                                                            
\bibitem[] { }
Dere, K.P. 1996, ApJ, 472, 864

\bibitem[] { }
Elbert, J.W., \& Sommers, P. 1995, \apj, 441, 151

\bibitem[] { }
Fryer, C.L., Benz, W., Herant, M., \& Colgate, S. 1999, astro-ph/9812058

\bibitem[] { }
Gallant, Y.A. \& Arons, J.  1994, \apj, 435, 230

\bibitem[] { }
Gosh, P., \& Lamb, F.K. 1978, \apj, 223, L83


\bibitem[]{}
Holman, G.D. 1985, ApJ, 293, 584

\bibitem[]{}
Jackson, J.D. 1981, in Classical Electrodynamics, p. 660

\bibitem[] { }
Kahler, S.W.,  1992, ARAA, 30, 113

\bibitem[] { }
Kim, \& Diamond 2000

\bibitem[]{}
LaRosa, T.N., Moore, R.L., Miller, J.A., \&
Shore, S.N. 1996, ApJ, 467, 454

\bibitem[] { }
Lazarian, A. \& Vishniac, E. 1999, \apj, , 517, 700

\bibitem[] { }
Lazarian, A. \& Vishniac, E. 2000, Rev. Mex. Astron. Astrof., 9, 55

\bibitem[] { }
Lawrence, M.A., Reid, R.J.O., \& Watson, A.A. 1991, J. Phys. Nucl. Part. Phys.,
17, 733

\bibitem[]{}
Litvinenko, Y.E. 1996, ApJ, 462, 997

\bibitem[]{}
Longair, M.S. 1992 in High Energy Astrophysics, vol. II, Chap. 21, p.


\bibitem[] { }
Medina Tanco, G.A., de Gouveia Dal Pino, E.M.,  \& Horvath, J. 1997, Astropart.
Phys., 6, 337

\bibitem[] { }
Medina Tanco, G.A., de Gouveia Dal Pino, E.M.,  \& Horvath, J. 1998, \apj, 492,
200


\bibitem[] { }
Olinto, A.V. 2000, Phys. Reports, 333, 329 

\bibitem[] { } 
Olinto, A.V. 2001, preprint (astro-ph/0102077)

\bibitem[] { }
Peebles, P.J.E. 1993, in Principles of Physical Cosmology, Princeton Univ. Press

\bibitem[] { }
Protheroe, R.J. 1999, in Topics in cosmic-ray astrophysics,  
ed. M. A. DuVernois (Nova Science Publishing:
     New York)  (astro-ph/9812055)

\bibitem[] { }
Protheroe, R.J., \& Johnson, P.A. 1995, Astropart. Phys., 4, 253

\bibitem[] { }
Rachen, J.P., \& Biermann, P.L. 1993, A\&A, 272, 161

\bibitem[] { }
Reames, D.V. 1995, Rev. Geophys., 33 (suppl.), 585

\bibitem[] { }
Sorrell, W.H. 1987, \apj, 323, 647

\bibitem[] { }
Shu, F.H., Najita, J., Ostriker, e., Wilkin, F., Ruden, S., and Lizano, S. 1994,
\apj, 429, 781

\bibitem[] { }
Shu, F.H. et al. 1998, in

\bibitem[] { }
Spruit, H.C.  1996, in NATO/ASI Ser. C417, Evolutionary Processes in Binary 
Stars (Dordrecht: Kluwer)

\bibitem[] { }
Stanev, T. 1997,\apj, 479, 290

\bibitem[] { }
Takeda, M. et al. 1999, \apj, 522, 225

\bibitem[] { }
Vishniac, E. , \& Lazarian, A. 1998, \apj, 511, 193
                                                

\bibitem[] { }
Woosley, S.E., \& Baron, E. 1992, \apj, 391, 228

\end{thebibliography}
\end{document}